\documentclass[twocolumn,showpacs,groupedaddress]{revtex4}
\usepackage{amssymb}
\usepackage{graphicx}
\usepackage{amsmath}
\begin{document}
\title{Single photon reflection and transmission  in optomechanical system}
\author{M.  A.  Khan\footnote{greatkhan17@gmail.com}, S. C. Hou, K. Farooq, and X. X. Yi\footnote{yixx@dlut.edu.cn}}
\affiliation{School of Physics and Optoelectronic Technology, Dalian
University of Technology, Dalian 116024, China}

\begin{abstract}
Cavity Optomechanical system is  speedily approaching the regime
where the radiation pressure of a single photon displaces the moving
mirror. In this paper, we consider a cavity optomechanical system
where the cavity field is driven by an external field. In the limit
of  weak   mirror-cavity couplings, we calculate analytically the
reflection and transmission rates for cavity field and discuss
the effects of mirror-cavity coupling on the reflection and
transmission.
\end{abstract}

 \maketitle

\section{INTRODUCTION}
Cavity Optomechanics  illustrates the radiation pressure by the
interaction between an optical cavity mode and the motion of a
mechanical object. The simplest example of such a system  is  a
Fabry-Perot cavity with a moveable mirror. Optomechanics is a
growing field of research studying the quantum dynamics of
electromagnetic and mechanical degree of freedom coupled through
radiation pressure and photothermal force or optical
gradient\cite{Daniel,Kipp,Marq}. In the fundamental optomechanical
setup, the frequency of an optical cavity modulates parametrically
with the position of mechanical oscillator. In most experiments,
this optomechanical coupling is small compared to mechanical
frequency and the linewidth of the cavity. However, if we drive an
optical cavity strongly, then the cavity may contain a large number
of photons, the coupling between cavity field and mechanical
oscillator would be increased by a factor $\sqrt{p}$, where $p$ is
the mean number of photon in the cavity. Recently,  this guides to
observe the radiation-pressure effect, for example, normal mode
splitting\cite{Grob}, optomechanically induced
transparency\cite{Weis,Safavi} and sideband
colling\cite{Gigan,Schliesser}.

The interaction of light with matter tells us a great  deal about
the nature of the matter. It covers variety of applications in
astrophysics, cosmology, quantum optics, and nanoscience. The
coupling between the  cavity field  and the  movable mirror in
optomechanics has been a great attraction for researchers because
they are helpful to  create  non-classical states of both cavity
field and mirror \cite{Bose}, and it has also been used for quantum
noise reduction\cite{Mancini}. Light-matter interactions can be
produced efficiently by using optical cavities adjustment of mirrors
that act as cavity resonators for light waves. Cavity quantum
electrodynamic(QED) system is significant  for examining light
matter interactions. Coupling of a single two-level atom with a
single mode of the electromagnetic field is reinforced by a cavity,
which is important for the investigation of light-matter
interaction\cite{ArkaMajumdar,Julsgaard, Scully, Dan, Albert}.
Jaynes-Cummings model  is the basic model for atom-field
interaction. This model consists of single mode radiation field
coupled with two level atoms. Due to presence of strong coupling,
the matter-field coupling increased the cavity field decay rate and
atomic decay rate\cite{Youn,Jaynes,Ficek}. The coupled atom-cavity
system can leads to a splitting in the atomic fluorescence spectrum
and the empty-cavity transmission resonances. This splitting is
known as vacuum-Rabi splitting\cite{Zhu,Sanchez, Agarwal}.
Vacuum-Rabi splitting can be detected dynamically in population
oscillation between two levels when field is resonant on the
transition\cite{Brune}, as well as in fluorescence spectrum when the
initial field strength is very small\cite{Sanchez} and in a cavity
transmission function profile at specific transition\cite{Zhu}. In
this work, we calculate analytical reflection and transmission
coefficients for field and intensities in the absence of phase noise
to prevent the coherence of the cavity field.

This paper is arranged  as follows. In section {\rm II},  we
describe the optomechanical system and its theoretical framework. In
section {\rm III}, we calculate steady state solution, leading
reflection and transmission coefficient for the field as well as
intensities. The findings are then discussed and presented in
Section {\rm IV}, while conclusion is given in section {\rm V}.

\begin{figure}
%\includegraphics[width=0.8\columnwidth,
%height=0.5\columnwidth] {TvsWl1.eps} \caption{
\includegraphics[width=0.7\columnwidth,
height=0.42\columnwidth]{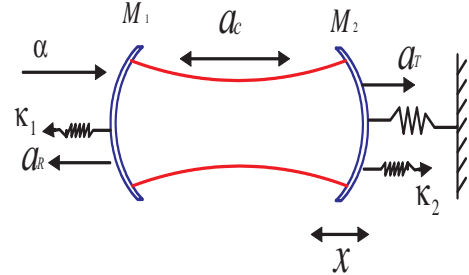} \caption{Schematic description of
an optomechanic  setup, which consists  of a mechanical oscillator
coupled to a cavity by radiation pressure. $\alpha$ is the input
field, $a_{c}$ the cavity field, and $a_R$ and $a_T$ is the
reflected and transmitted field, receptively. $\kappa_1$ and
$\kappa_2$ denotes the cavity loss rate  via the left and right
mirror, respectively.}
\end{figure}

\section{Theoretical framework}

We consider an optomechanical cavity with frequency $\omega_c$
formed by a fixed end mirror $M_1$ and a moving end mirror $M_2$.
The mirror $M_2$ can be considered as a harmonic oscillator with
mass $m$ and frequency $\omega _m$. The oscillator mirror and the
cavity are coupled with each other via radiation pressure. The
system is coherently driven by the field $\alpha=\alpha
e^{-i\omega_{L} t}$ with frequency $\omega_{L}$ as shown in figure
1. The cavity decay rate of mirrors are denoted by $\kappa_{1}$ and
$\kappa_{2}$. In the rotating frame with driving frequency
$\omega_{L}$, the total
Hamiltonian\cite{Julsgaard,Kumar,Man'ko,Jacobs,Jie} of the system
can be written as
\begin{equation}
\begin{aligned}
H= \hbar \Delta_{c} \hat{a}_c^\dagger \hat{a}_c+
\frac{\hat{p}^2}{2m}+\frac{1}{2}m \omega_{m}^2 \hat{x}^2
-\hbar g \hat{a}_c^\dagger \hat{a}_c \hat{x}\\
+i\hbar\sqrt{2\kappa_{1}}( \alpha\hat{a}_c^\dagger-\alpha^*
\hat{a}_c),
\end{aligned}
\end{equation}
Where $\Delta_{c}=\omega_c-\omega_L$ is the cavity resonance
frequency detuning. $\hat{a}_c$ and $\hat{a}_c^\dagger $ are,
respectively, the annihilation and creation operators for the cavity
field of frequency $\omega_c$ with the commutation relation
$[\hat{a}_c,\hat{a}_c^\dagger]=1$. $\hat{p}$ and $\hat{x}$ are
the momentum and displacement of the oscillating mirror with mass
$m$ and frequency $\omega_{m}$. The parameter $g$ is  coupling
strength of the radiation pressure between the mirror and the
cavity, which is considered sufficiently weak in this paper.

The  input-output theory for a light field interacting with a cavity
is given by\cite{collett}
\begin{equation}
\begin{aligned}
\hat{a}_R=\sqrt{2\kappa_{1}}\hat{a}_c-\alpha\\
\hat{a}_T=\sqrt{2\kappa_{2}}\hat{a}_c\\
\end{aligned}
\end{equation}
The  first and second equations  determine the reflected and
transmitted field of the input and output mirror, using the cavity
decay rate $\kappa_{1}$ and $\kappa_{2}.$

The dynamics of the the system is determined by using the master equation
\begin{equation}
\begin{aligned}
\frac{\partial \rho}{\partial
t}=\frac{1}{i\hbar}[H,\rho]+\mathcal{L}(\rho),
\end{aligned}
\end{equation}
where $\mathcal{L}(\rho)$ is  the Lindblad operator. This is used to
model the incoherent decay processes and is given by
\begin{equation}
\begin{aligned}
\mathcal{L}(\rho)=D\rho D^\dagger-\frac{1}{2}\rho D^\dagger D-\frac{1}{2}D^\dagger D \rho
\end{aligned}
\end{equation}
The cavity mode is damped by  photon leakage, which is designed by
Lindblad term with $D=\sqrt{2\kappa}\hat{a}_c$, where $\kappa$ is
the total cavity-field decay rate and is given by
$\kappa=\kappa_{1}+\kappa_{2}$

By using the master equation, the dynamics of the system can be
written as
\begin{equation}
\begin{aligned}
\frac{\partial \langle \hat{a}_c\rangle}{\partial t}
=-i\Delta_{c}\langle \hat{a}_c\rangle +ig\langle
\hat{x}\rangle\langle
\hat{a}_c\rangle+\sqrt{2\kappa_{1}}\alpha-\kappa\langle
\hat{a}_c\rangle
\end{aligned}
\end{equation}
\begin{equation}
\begin{aligned}
\frac{\partial \langle
\hat{a}_c^\dagger\hat{a}_c\rangle}{\partial t} =-2\kappa\langle
\hat{a}_c^\dagger\hat{a}_c\rangle+\sqrt{2\kappa_{1}}(
\alpha\langle\hat{a}_c^\dagger\rangle+\alpha^*
\langle\hat{a}_c\rangle)
\end{aligned}
\end{equation}
\begin{equation}
\begin{aligned}
\frac{\partial \langle \hat{x}\rangle}{\partial t}=\frac{1}{m}\langle \hat{p}\rangle
\end{aligned}
\end{equation}
\begin{equation}
\begin{aligned}
\frac{\partial \langle \hat{p}\rangle}{\partial t}
=-m\omega_{m}^2\langle \hat{x}\rangle+\hbar g\langle
\hat{a}_c^\dagger\hat{a}_c\rangle-\frac{\gamma}{m}\langle \hat{p}\rangle
\end{aligned}
\end{equation}

Where $\gamma$ is  damping rate of the mechanical
oscillator.
\begin{figure}
\includegraphics[width=0.7\columnwidth,
height=0.42\columnwidth] {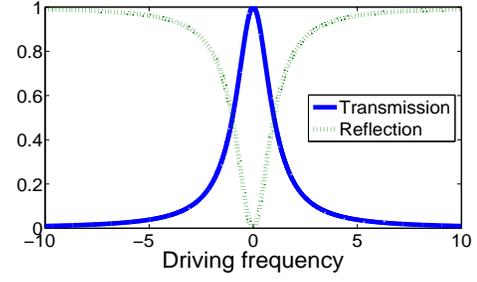} \caption{The
cavity-intensity-transmission and reflection coefficient as a
function of driving frequency. Here $\kappa_{1}=\kappa_{2}=0.5$, $m=1$,
$\omega_m=1$, $\omega_c=0$, $\gamma=0$, $|\alpha|^2=1$, $g=0.1$ and $\hbar=1$ .}
\end{figure}

\section{Steady State solution}

The steady state solution of equations (5-8) can be worked out by
putting the time derivative equal to zero. The steady state solution
of these equations leads to
\begin{equation}
\begin{aligned}
\langle \hat{a}_c\rangle=\frac{\sqrt{2\kappa_{1}}\alpha}
{\kappa+i(\Delta_{c}-g\langle \hat{x}\rangle)}
\end{aligned}
\end{equation}
\begin{equation}
\begin{aligned}
\langle \hat{a}_c^\dagger\hat{a}_c\rangle=|\langle \hat{a}_c\rangle|^2
\end{aligned}
\end{equation}
\begin{equation}
\begin{aligned}
\langle \hat{p}\rangle=0
\end{aligned}
\end{equation}
\begin{equation}
\begin{aligned}
\langle \hat{x}\rangle=X_{1}-X_{2}+X_{3}.
\end{aligned}
\end{equation}

Here $X_{1}=\frac{2\Delta_{c}}{3g}$,
$X_{2}=\frac{2^{\frac{1}{3}}m_{1}}{3g^2m\omega_{m}^2(m_{2}
+\sqrt{4m_{1}^3+m_{2}^2})^{\frac{1}{3}}}$,
$X_{3}=\frac{(m_{2}+\sqrt{4m_{1}^3+m_{2}^2})^{\frac{1}{3}}}{
2^{\frac{1}{3}}3g^2m\omega_{m}^2}$, $m_{1}=3g^2\kappa^2m^2
\omega_{m}^4-g^2m^2\omega_{m}^4\Delta_{c}^2$ and
$m_{2}=54g^5m^2\omega_{m}^4\alpha\hbar\alpha^*\kappa_{1}
-18g^3\kappa^2m^3\omega_{m}^6\Delta_{c}-2g^3m^3\omega_{m}^6\Delta_{c}^3$
$\langle \hat{a_{c}}\rangle$  and $\langle
\hat{a_{c}}^\dagger\hat{a_{c}}\rangle$ are field amplitude and
photon number. These results are very important for the manipulation
of transmission  and reflection  intensity of the filed.  The
complex field-transmission and reflection coefficients can be
written as
\begin{equation}
\begin{aligned}
t=\frac{\langle \hat{a}_T\rangle}{\alpha}
=\frac{2\sqrt{\kappa_{1}\kappa_{2}}}{\kappa+i(\Delta_{c}-g\langle
\hat{x}\rangle)}
\end{aligned}
\end{equation}

\begin{equation}
\begin{aligned}
r=\frac{\langle \hat{a}_R\rangle}{\alpha}
=\frac{2\kappa_{1}}{\kappa+i(\Delta_{c}-g\langle\hat{x}\rangle)}-1
\end{aligned}
\end{equation}

Intensity-reflection coefficient $R$,
\begin{equation}
\begin{aligned}
R=\frac{ \langle \hat{a}_R^\dagger\hat{a}_R\rangle
}{|\alpha|^2}.
\end{aligned}
\end{equation}
Intensity-transmission coefficient $T$,
\begin{equation}
\begin{aligned}
T=\frac{\langle \hat{a}_T^\dagger\hat{a}_T\rangle }{|\alpha|^2}
\end{aligned}
\end{equation}
By using equation (2), we can find $R$ and $T$.
\begin{equation}
\begin{aligned}
R=|r|^2=\frac{4\kappa^2_{1}} {\kappa^2+(\Delta_{c}
-g\langle\hat{x}\rangle)^2}
-\frac{2\kappa_{1}}{\kappa+i(\Delta_{c}-g\langle\hat{x}\rangle)}\\
-\frac{2\kappa_{1}}{\kappa-i(\Delta_{c}-g\langle\hat{x}\rangle)}+1
\end{aligned}
\end{equation}

\begin{equation}
\begin{aligned}
T=|t|^2=\frac{4\kappa_{1}\kappa_{2}}{\kappa^2+(\Delta_{c}-g\langle\hat{x}\rangle)^2}
\end{aligned}
\end{equation}
Here $r$ and $t$, respectively,  are the coefficient of reflection
and transmission for the fields given by equation (13,14).

\begin{figure}
\includegraphics[width=0.7\columnwidth,
height=0.42\columnwidth] {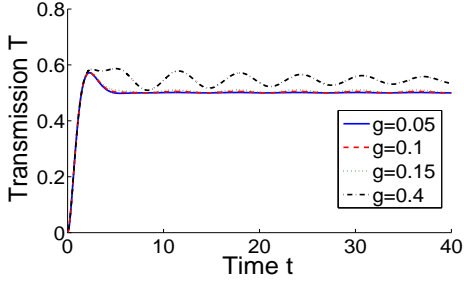} \caption{Dynamics of the
cavity-intensity-transmission  coefficient T. Here
$\kappa_{1}=\kappa_{2}=0.5$, $m=1$, $\omega_m=1$, $\gamma=0$,
$\Delta_{c}=1$, $|\alpha|^2=1$ and $\hbar=1$. Initial condition for
given curve  is $\langle \hat{a}_c\rangle=\langle
\hat{a}_c^\dagger\hat{a}_c\rangle=\langle
\hat{p}\rangle=\langle\hat{x}\rangle=0$}
\end{figure}

\section{Results and discussions}

Equations (17) and (18) are the main result of the present paper.
The sum of $R$ and $T$ is equal to 1. The cavity intensity
transmission and reflection coefficient as a function of driving
frequency are shown in Fig.2. When detuning is zero, the
transmission rate arrives at its maximum. As we increased or
decreased the detuning, the  transmission rate becomes smaller and
smaller. In our work, we use an  approximation,  i.e., $\langle
\hat{x}\hat{a}_c\rangle=\langle \hat{x}\rangle\langle
\hat{a}_c\rangle$.

\begin{figure}
\includegraphics[width=1.0\columnwidth,
height=0.43\columnwidth] {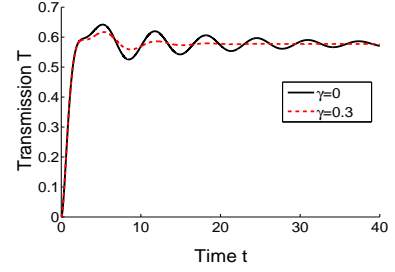} \caption{Dynamics of the
cavity-intensity-transmission  coefficient with undamped and damped of the movable mirror. Here
$\kappa_{1}=\kappa_{2}=0.5$, $m=1$, $\omega_m=1$, $\Delta_{c}=1$, $|\alpha|^2=1$, $g=0.5$
and $\hbar=1$. Initial condition for given curve is $\langle \hat{a}_c\rangle=\langle \hat{a}_c^\dagger\hat{a}_c\rangle=\langle \hat{p}\rangle=\langle\hat{x}\rangle=0$}
\end{figure}

The  condition which ensures the validity of the weak
coupling approximation  can be derived from Eq.(5-8). To make the
derivation clear, we   ignore the damping of the harmonic
oscillator, i.e., $\gamma=0.$  The weak coupling condition can be
found by examining  the linear dependence of   $\langle
\hat{a}_c\rangle$ on the coupling constant $g$. For this purpose, we
ignore the term with $g$ in equation (8) and have $\frac{\partial^2
\langle \hat{x}\rangle}{\partial t^2}=-\omega_{m}^2\langle
\hat{x}\rangle$, leading to $\langle
\hat{x}\rangle=x_0\sin(\omega_{m}t+\phi)$, where $x_0=\langle
\hat{x}(t=0)\rangle$ is the maximum amplitude of oscillation and
$\phi$ is the initial phase angle. Substituting this result  into
equation (5), we have
\begin{equation}
\begin{aligned}
\frac{\partial \langle \hat{a}_c\rangle}{\partial t}=
-i\Delta_{c}\langle \hat{a}_c\rangle+ig\langle \hat{a}_c\rangle x_0\sin(\omega_{m}t+\phi)\\
+\sqrt{2\kappa_{1}}\alpha-\kappa\langle \hat{a}_c\rangle.
\end{aligned}
\end{equation}
The resulting $\langle\hat{a}_c\rangle$ clearly is a function of
time $t$ and $g$, i.e.,
$\langle\hat{a}_c\rangle=\langle\hat{a}_c(g,t)\rangle$. To find how
$\langle\hat{a}_c(g,t)\rangle$ linearly depends on $g$, we expand
$\langle\hat{a}_c\rangle$ as follows,
\begin{equation}
\begin{aligned}
\langle \hat{a}_c(g,t)\rangle=\langle\hat{a}_c(0,t)\rangle
+g\langle \hat{a}_{c1}(t)\rangle+g^2\langle \hat{a}_{c2}(t)\rangle+...
\end{aligned}
\end{equation}
Substituting this expansion into Eq. (19), we find for the zeroth
order of $g$,
\begin{equation}
\begin{aligned}
\frac{\partial \langle \hat{a}_c(0,t)\rangle}{\partial t}=-i\Delta_{c}\langle \hat{a}_c\rangle+\sqrt{2\kappa_{1}}\alpha-\kappa\langle \hat{a}_c\rangle
\end{aligned}
\end{equation}
For first order of $g$, we have
\begin{equation}
\begin{aligned}
\frac{\partial \langle \hat{a}_{c1}(t)\rangle}{\partial t}=(-i\Delta_{c}-\kappa)\langle \hat{a}_{c1}(t)\rangle\\
+i\langle \hat{a}_c(0,t)\rangle x_0\sin(\omega_{m}t+\phi)
\end{aligned}
\end{equation}
After taking integration of equation (22) and putting the result into equation (20), we have
\begin{equation}
\begin{aligned}
\langle \hat{a}_c(g,t)\rangle=\langle \hat{a}_c(0,t)\rangle+\frac{g x_0}{2 \omega_{m}}\langle \hat{a}_c(0,t)\rangle \sin(\omega_{m}t+\phi)
\end{aligned}
\end{equation}
Clearly, the weak coupling limit means,
\begin{equation}
\begin{aligned}
\eta=\frac{g x_0}{2\omega_m}\ll1
\end{aligned}
\end{equation}
It is the condition for the validity of weak coupling regime. Fig.3
shows the dynamics of the intensity-transmission coefficient $T$ as
a function of time. For small value of coupling constant, our
approximation is valid,i.e. the transmission can reach a steady
value as time approaches infinity. As we increased the value of
coupling constant from $g=0.1$ to $g=0.4$,  there is a  large
oscillation in transmission $T$, which means  our approximation is
not valid  for strong couplings. However, If we use damping of
movable mirror, then we may consider the strong coupling, because
the damping term speed up damping and our system comes to a
stationary state very quickly as shown in Fig.4. For steady state
result, we plot the ratio of transmission rate at $g=0$ to that of
$g\neq 0$, i.e. $\frac{T(g\neq0)}{T(g=0)})$, see Fig.5. It shows
that, for large detuning, transmission rates are independent of
coupling constant and their ratio ia almost 1.  For small detuning,
however, the coupling constant has very large effect on the
transmission rate. The larger the value of coupling constant, the
larger the transmission rate is. For positive detuning,  i.e.
$\Delta_c>0$, transmission rate with coupling   is larger than the
transmission rate without coupling $T(g \neq 0)>T(g=0)$. For
negative detuning, i.e. $\Delta_c<0$, transmission rate with
coupling   is smaller than the transmission rate without
coupling,i.e. $T(g\neq0)<T(g=0)$.

\begin{figure}
\includegraphics[width=0.7\columnwidth,
height=0.45\columnwidth] {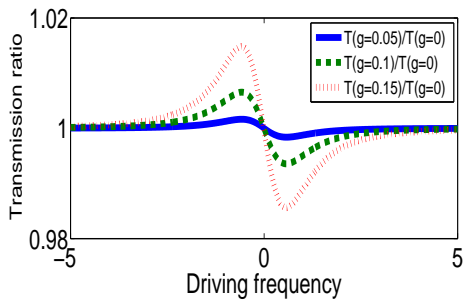} \caption{The
cavity-intensity-transmission ratio $\frac{T(g\neq0)}{T(g=0)})$ as a
function of driving frequency $\omega_L$. For given  curves,
$\kappa_{1}=\kappa_{2}=0.5$, $m=1$, $\omega_m=1$, $\omega_c=0$, $\gamma=0$, $|\alpha|^2=1$ and
$\hbar=1$ .}
\end{figure}

\section{Conclusions}
In this paper, we study the transmission and reflection rates of an
optical cavity, which has a moving mirror and is driven coherently
by an external field. We aim at the effects of the moving mirror on
the transmission and reflection. Our analysis is based on the weak
mechanical interaction. We calculated analytically the steady state
transmission and reflection coefficients as a function of detuning
$\Delta_c$ and of the cavity-mirror coupling constant. In addition,
we show that for small value of coupling constant $g$, our
approximation is valid, but it fails to treat the case of strong
couplings.

\

%\begin{acknowledgements}
M. A. Khan and K. Farooq acknowledge China Scholarship Council(CSC)
for the Research Fellowship. Hou and Yi acknowledge  the financial
support by the NSF of China under Grants Nos 61078011, 10935010 and
11175032.
%\end{acknowledgements}

\end{document}